\documentstyle[psfig,mncite,subfigure]{mn}

\begin{document}

\title{Prospects for Spectroscopic Reflected-light Planet Searches}
\author
[C.Leigh, A.Collier~Cameron, T.Guillot]
{\Large Christopher Leigh $^{1}$, Andrew Collier Cameron $^{1}$, Tristan Guillot $^{2}$ \\
\small $^{1}$ University of St Andrews, St Andrews, Fife, KY16 9SS, U.K \\
$^{2}$ Observatoire de la Cote d'Azur, B.P. 4229, 06304 Nice, Cedex 4, France}
\date{\today}

\maketitle

\begin{abstract}
High-resolution spectroscopic searches for the starlight reflected from
close-in extra-solar giant planets have the capability of determining the 
optical albedo spectra and scattering properties of these objects. When combined
with radial velocity measurements they also yield the true mass of the planet.
To date, only two such planets have been targeted for reflected-light signals,
yielding upper limits on the planets' optical albedos. Here we examine the prospects
for future searches of this kind. We present Monte Carlo estimates of prior
probability distributions for the orbital velocity amplitudes and planet/star flux 
ratios of 6 bright stars known to harbour giant planets in orbits with 
periods of less than 5 days. Using these estimates, we assess the viability of 
these targets for future reflected-light searches using 4m and 8m-class telescopes.
\\
\\
\\
\end{abstract}
\section{Introduction}

Spectral models of close-orbiting giant exoplanets 
\cite{seager98,marley99,gouken99,sudarsky2000,barman01,sudarsky03,baraffe03}
show that scattered starlight dominates over thermal emission at optical wavelengths, but the
albedo is very sensitive to the depth of cloud formation. \scite{sudarsky2000} find that the
relatively low surface gravities of planets such as $\upsilon$ Andromeda b, HD 75289 b and HD
209458 b may favour the formation of relatively bright, high-altitude silicate cloud decks. On
higher-gravity objects such as $\tau$ Bootis b, however, the same models predict a much deeper
cloud deck. In this case, much of the optical spectrum is absorbed by the resonance lines of the
alkali metals. The Na I D lines in particular are strongly broadened by collisions with $H_{2}$
molecules.

In reality, however, the atmospheric structure of close-orbiting giant planets
(hereafter ``Pegasi Planets'') is expected to be more complex, with strong winds
and horizontal temperature variations \cite{showman02}, and possible temporal
variations \cite{cho03}. The combined atmospheric circulation and
temperature variations are likely to yield a disequilibrium chemical composition.
Furthermore, the location and characteristics of any cloud decks cannot
be predicted to any great precision with present theoretical knowledge.
The next logical step towards understanding the properties and evolution of
these objects will be to test these emerging models by measuring the albedos of
these Pegasi planets directly.

\pagebreak

A planet of radius $R_{p}$ orbiting a distance $a$ from a star intercepts a fraction $\left
(R_{p} / 2a \right)^{2}$ of the stellar luminosity. The proximity of Pegasi planets to their
parent stars make them excellent prospects for reflected-light searches. At optical
wavelengths, starlight scattered off the planet's atmosphere is expected to dominate over
thermal emission. \scite{charb99,cameron99} showed that the planet/star flux ratio
\begin{equation}
\varepsilon(\alpha,\lambda) \equiv \frac{f_{p}(\alpha,\lambda)}{f_{\star}(\lambda)}
   = p(\lambda) \left( \frac{R_{p}}{a} \right) ^{2} g \left( \alpha,\lambda \right)
\label{eq:epsilon}
\end{equation}
is expected to vary with the star-planet-observer illumination angle $\alpha$, giving
a small periodic modulation to the system brightness in the form of the phase function $g
(\alpha,\lambda)$. Here $p(\lambda)$ represents the geometric albedo of
the planet's atmosphere that, like the phase function, is wavelength ($\lambda$) dependent.
Assuming a grey albedo model, i.e. no wavelength dependence, the amplitude of optical flux
variability for a typical Pegasi planet system is expected to be
\begin{equation}
\frac{\Delta f}{f}
     \leq 8.3 \times 10^{-5} \left (\frac{p}{0.4} \right)
     \left(\frac{R_{p}/1.4R_{Jup}}{a/0.045AU} \right) ^{2}
\label{eq:fluxratio}
\end{equation}
Future space-based photometry missions are expected to be able to detect and measure this
modulation with ease. However, the orbital inclination (and hence planet's mass) cannot be
determined directly form the light-curve, since the form of the phase function $g
(\alpha,\lambda)$ is unknown.

\section{Spectroscopic reflected-light searches}

\begin{table*}
\begin{tabular}{cccccccccc}
\hline \hline \\
Star & Spectral & $m_{V}$ & $T_{eff}$ & $P_{rot}$     & $v \sin{i}$   & [Fe/H] & Age     & $M_{*}/M_{\odot}$ & $R_{*}/R_{\odot}$ \\
     & Type     & (Mags)  & (K)       & (Days)        & $(km.s^{-1})$ &        & (Gyrs)  &                   &                   \\
\hline
$\upsilon$ And & F8V $^{1}$ & 4.09  & 6107$\pm$80 $^{4}$  & 12.2$\pm$2  $^{6}$  & 9.3$\pm$0.4  $^{4}$  & 0.09$\pm$0.06  $^{4}$  & 3.8$\pm$1.0 $^{4}$
               & 1.27$\pm$0.06 $^{14}$ & 1.67$\pm$0.06 $^{14}$ \\
$\tau$ Boo     & F7V $^{1}$ & 4.50  & 6360$\pm$80 $^{4}$  & 3.3$\pm$0.1 $^{7}$  & 14.8$\pm$0.3 $^{4}$  & 0.27$\pm$0.08  $^{4}$  & 1.0$\pm$0.6 $^{4}$
               & 1.37$\pm$0.05 $^{14}$ & 1.46$\pm$0.06 $^{14}$ \\ 
51 Peg         & G2V $^{1}$ & 5.46  & 5793$\pm$70 $^{4}$  & 29$\pm$7    $^{6}$  & 2.1$\pm$0.7  $^{4}$  & 0.20$\pm$0.07  $^{4}$  & 4.0$\pm$2.5 $^{4}$
               & 1.10$\pm$0.06 $^{14}$ & 1.20$\pm$0.07 $^{14}$ \\
HD 179949      & F8V $^{2}$ & 6.25  & 6115$\pm$50 $^{5}$  & 9$\pm$2     $^{8}$  & 6.3$\pm$0.9  $^{9}$  & 0.22$\pm$0.07  $^{11}$ & 3.5$\pm$2.5 $^{13}$
               & 1.24$\pm$0.10 $^{14}$ & 1.24$\pm$0.10 $^{14}$ \\
HD 75289       & G0V $^{3}$ & 6.35  & 6000$\pm$50 $^{3}$  & 16$\pm$3    $^{8}$  & 4.4$\pm$1.0  $^{10}$ & 0.29$\pm$0.08  $^{12}$ & 5.6$\pm$1.0 $^{12}$
               & 1.22$\pm$0.05 $^{14}$ & 1.25$\pm$0.05 $^{14}$ \\
HD 209458      & F9V $^{1}$ & 7.65  & 6000$\pm$50 $^{8}$  & 16.5$\pm$3  $^{18}$ & 3.7$\pm$1.3  $^{17}$  & 0.00$\pm$0.07  $^{15}$ & 5.7$\pm$1.0 $^{16}$
               & 1.10$\pm$0.10 $^{15}$ & 1.20$\pm$0.10 $^{15}$ \\
\hline \hline \\
\end{tabular}
\caption[]{Stellar parameters for bright stars known to harbour Pegasi planets. $^{1}$ \scite{gray01}, $^{2}$ \scite{houk88},
$^{3}$ \scite{gratton89}, $^{4}$\scite{fuhrmann98}, $^{5}$ \scite{gray92}, $^{6}$ \scite{henry2000},
$^{7}$ Synchronised rotation assumed from \scite{baliunas97,henry2000},
$^{8}$ \scite{mazeh2000}, $^{9}$ \scite{groot96},
$^{10}$ \scite{benz84}, $^{11}$ \scite{tinney01}, $^{12}$ \scite{udry2000}, $^{13}$ Estimate based on spectral type
$^{14}$ Average of Spectroscopic studies by \scite{fuhrmann97,fuhrmann98,gonzalez98,gonzalez2000,gonzalez01} 
$^{15}$ Estimated from \scite{tinney01,udry2000}, $^{16}$ \scite{cody02}
$^{17}$ \scite{queloz2000}, $^{18}$ Average from \scite{barnes01,queloz2000}
}
\label{stellarparam}
\end{table*}

\begin{table*}
\begin{tabular}{cccccccc}
\hline \hline \\
Planet & $K_{*}$       & $P_{orb}$ & Orbital     & $M_{p}/M_{Jup}~ \sin{i}$  & $T_{Eff}$ & $R_{p}/R_{Jupiter}$ & $R_{p}/R_{Jupiter}$      \\
     & $(m.s^{-1})$  & (Days)    & Radius (au) &                           & (K)       & Hot model		 & Cold Model          \\
\hline \\
$\upsilon$ And b & 74.5$\pm$2.3 $^{1}$ & 4.6171$\pm$0.0001 $^{1}$  & 0.0588$\pm$0.0020 & 0.716$\pm$0.053 & 1570 & 1.49 & 1.32  \\
$\tau$ Boo b     & 469$\pm$5    $^{2}$ & 3.3125$\pm$0.0002 $^{1}$  & 0.0483$\pm$0.0018 & 4.242$\pm$0.224 & 1680 & 1.37 & 1.16  \\
51 Peg b         & 56$\pm$1     $^{3}$ & 4.2306$\pm$0.0005 $^{3}$  & 0.0528$\pm$0.0029 & 0.475$\pm$0.038 & 1330 & 1.53 & 1.48  \\
HD 179949 b      & 102$\pm$3    $^{4}$ & 3.0930$\pm$0.0001 $^{4}$  & 0.0446$\pm$0.0036 & 0.838$\pm$0.109 & 1550 & 1.40 & 1.31  \\
HD 75289 b       & 54$\pm$1     $^{5}$ & 3.5098$\pm$0.0007 $^{5}$  & 0.0483$\pm$0.0020 & 0.461$\pm$0.028 & 1470 & 1.58 & 1.70  \\
HD 209458 b      & 81$\pm$6     $^{6}$ & 3.5239$\pm$0.0001 $^{6}$  & 0.0470$\pm$0.0018 & 0.620$\pm$0.050 & 1460 & 1.38 & 1.36  \\
\hline \hline \\
\end{tabular}
\caption[]{System parameters for known Pegasi planets around bright stars. $^{1}$ Private Communication - Geoff Marcy, $^{2}$
\scite{butler97}, $^{3}$ \scite{mayor95,marcy97}, $^{4}$ \scite{tinney01}, $^{5}$ \scite{udry2000}, $^{6}$ \scite{henry2000a}.
The results in columns 4 and 5 are calculated from stellar and planetary parameters using the equations detailed in
Section~\ref{sec:parameters}. Effective planetary temperature and radius estimates are given for the estimated age of
the system, assuming an edge-on inclination $i=90$ (i.e. upper limit).
}
\label{planetparam}
\end{table*}

Starlight reflected from a planet's atmosphere contains copies of all the thousands
of photospheric stellar absorption lines, Doppler-shifted by the planet's orbital
motion and modulated in strength by the phase function $g(\alpha,\lambda)$.
By detecting and characterising the planetary reflected-light signature we
observe the planet/star flux ratio as a function of orbital phase and
wavelength. The information we aim to obtain (in order of increasing
difficulty) comprises:

\smallskip
\noindent $\bullet~K_{p}$, the planet's projected orbital velocity. From this we
learn the orbital inclination $i$ and hence the planet mass $M_{p}$, since 
$M_{p} \sin{i}$ is known form the star's Doppler wobble.

\smallskip
\noindent $\bullet~\epsilon_{0}$, the maximum strength of the reflected starlight that 
would be seen if the planet were viewed fully illuminated.

\smallskip
\noindent $\bullet~p(\lambda)$, the albedo spectrum, which depends on the composition 
and structure of the planet's atmosphere.

\smallskip
\noindent $\bullet~g(\alpha,\lambda)$, the phase function describing the dependence of 
the amount of light reflected toward the observer by the star-planet-observer angle
$\alpha$.

\subsection{Current Status}

%\smallskip
\noindent Several attempts have been made to detect this ``faint'' echo of the
starlight, using high-precision spectral subtraction and pattern-matching
techniques. \scite{charb99,cameron99} and \scite{leigh03} searched independently for
starlight reflected from $\tau$ Bootis $b$, establishing an upper limit on the 
planet/star flux ratio that would be seen if the planet were viewed fully illuminated,
with
\begin{equation}
\varepsilon_{0} = p \left( \frac{R_{p}}{a} \right) ^{2} < 10^{-4}
\label{eq:varepsilon}
\end{equation}
\noindent for the observed wavelength range of 407 - 649 nm.  For an assumed planetary radius
$R_{p}=1.2~R_{Jup}$, the \scite{leigh03} 99.9\% confidence levels (Fig. \ref{fig:limits}a)  
suggest a geometric albedo $p<0.39$ at the most probable orbital inclination $i \sim 40
^{\circ}$, assuming a grey albedo model.

\scite{cameron02} used the same technique on $\upsilon$ Andromeda $b$, establishing an upper
radius limit $R_{p} < 1.22 R_{Jup}$, assuming a grey geometric albedo of $p=0.5$.
(Fig. \ref{fig:limits}b). Both \scite{cameron02} and \scite{leigh03} find possible detections
above the $2 \sigma$ significance level, which if confirmed would yield masses of 0.74 and
7.28 $M_{Jup}$ respectively for $\upsilon$ And $b$ and $\tau$ Bootis $b$. However, there is a
$\sim$ 4 to 8\% probability that the detected features, shown at Figure \ref{fig:limits}, are a consequence
of non-gaussian noise, and as a result neither are confident enough to claim a genuine detection.

\section{System Parameters}
\label{sec:parameters}

\begin{figure*}
\begin{center}
\begin{tabular}{cc}
\psfig{figure=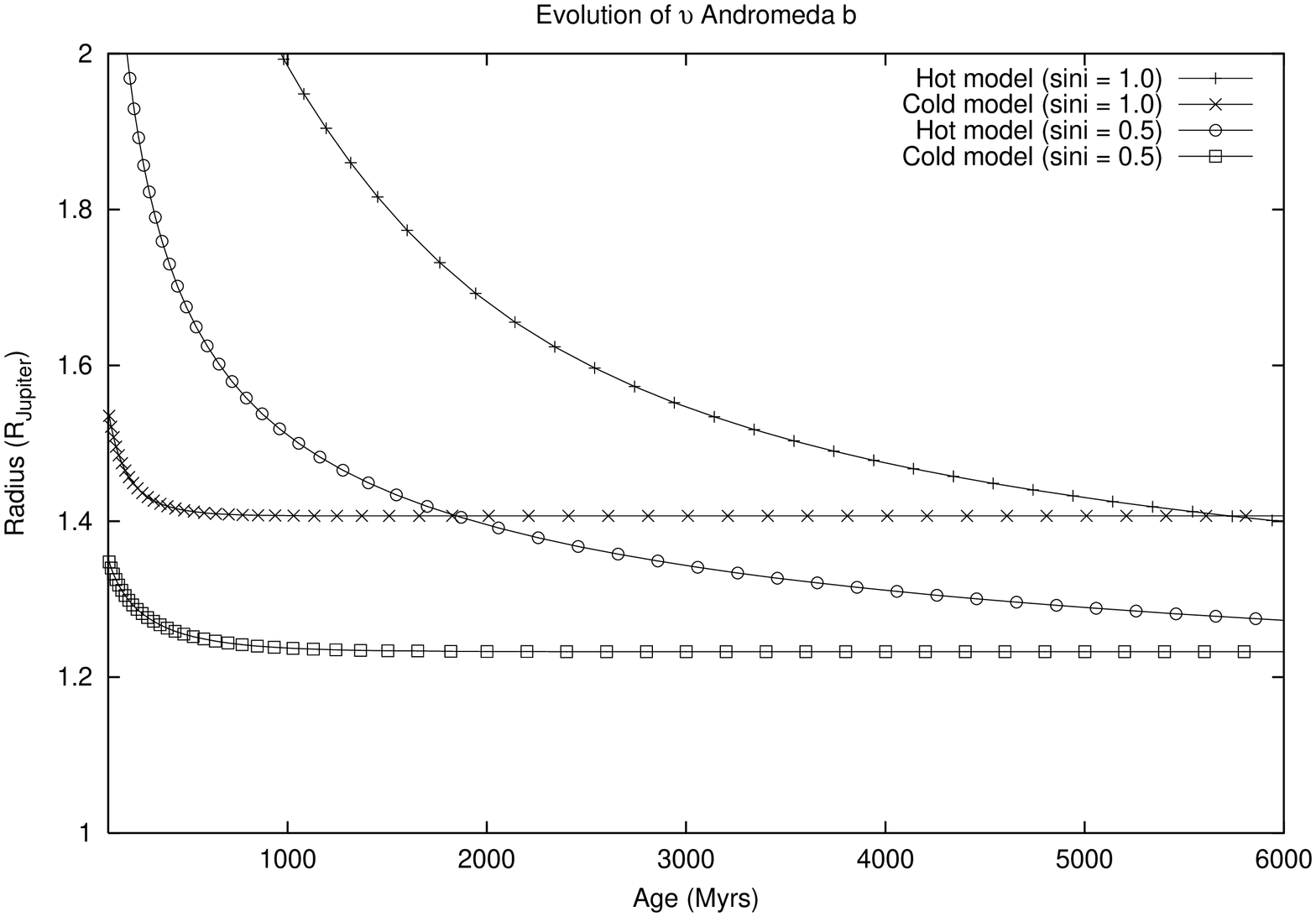,angle=-90,width=8cm} &
\psfig{figure=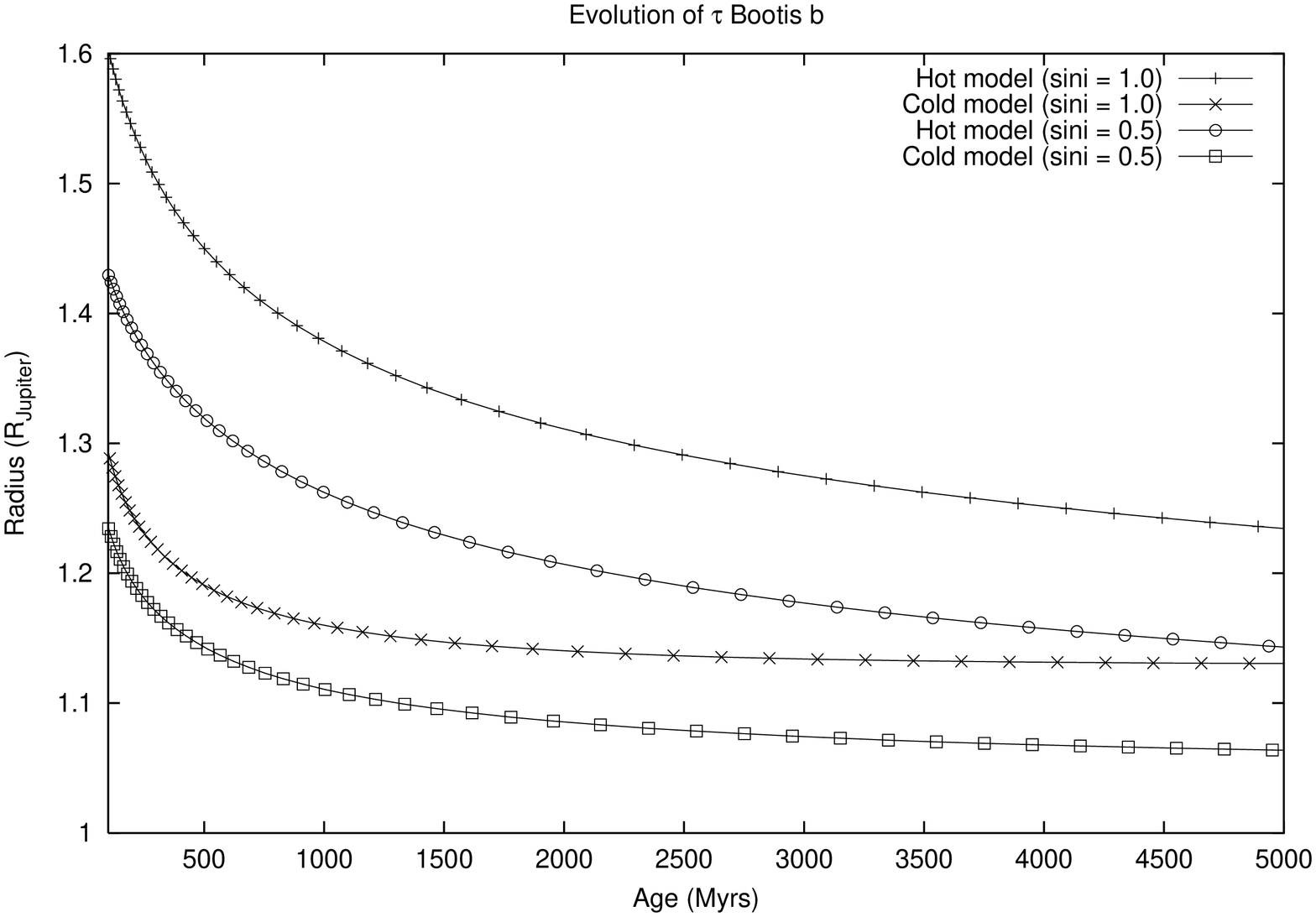,angle=-90,width=8cm} \\
\psfig{figure=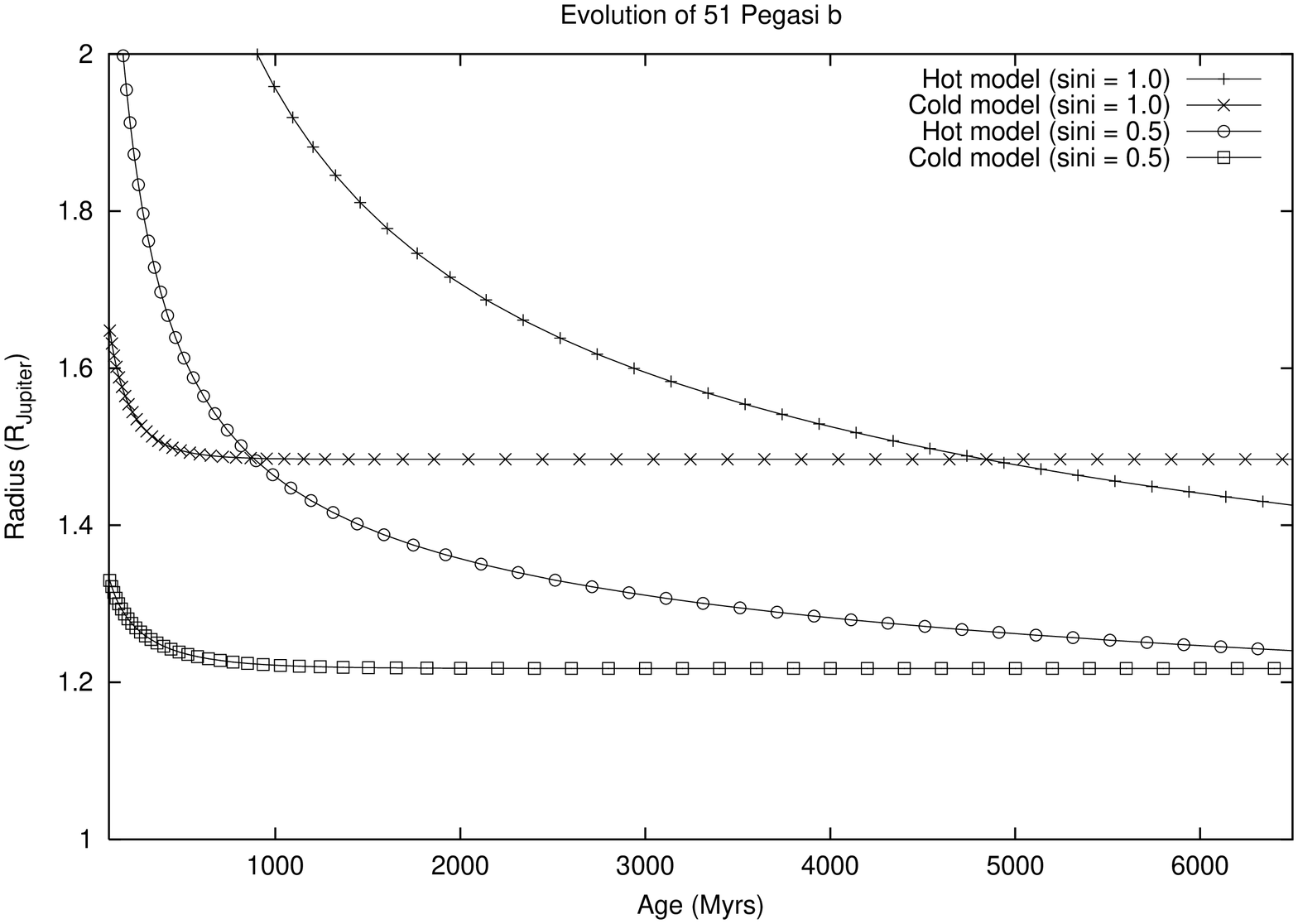,angle=-90,width=8cm}  &
\psfig{figure=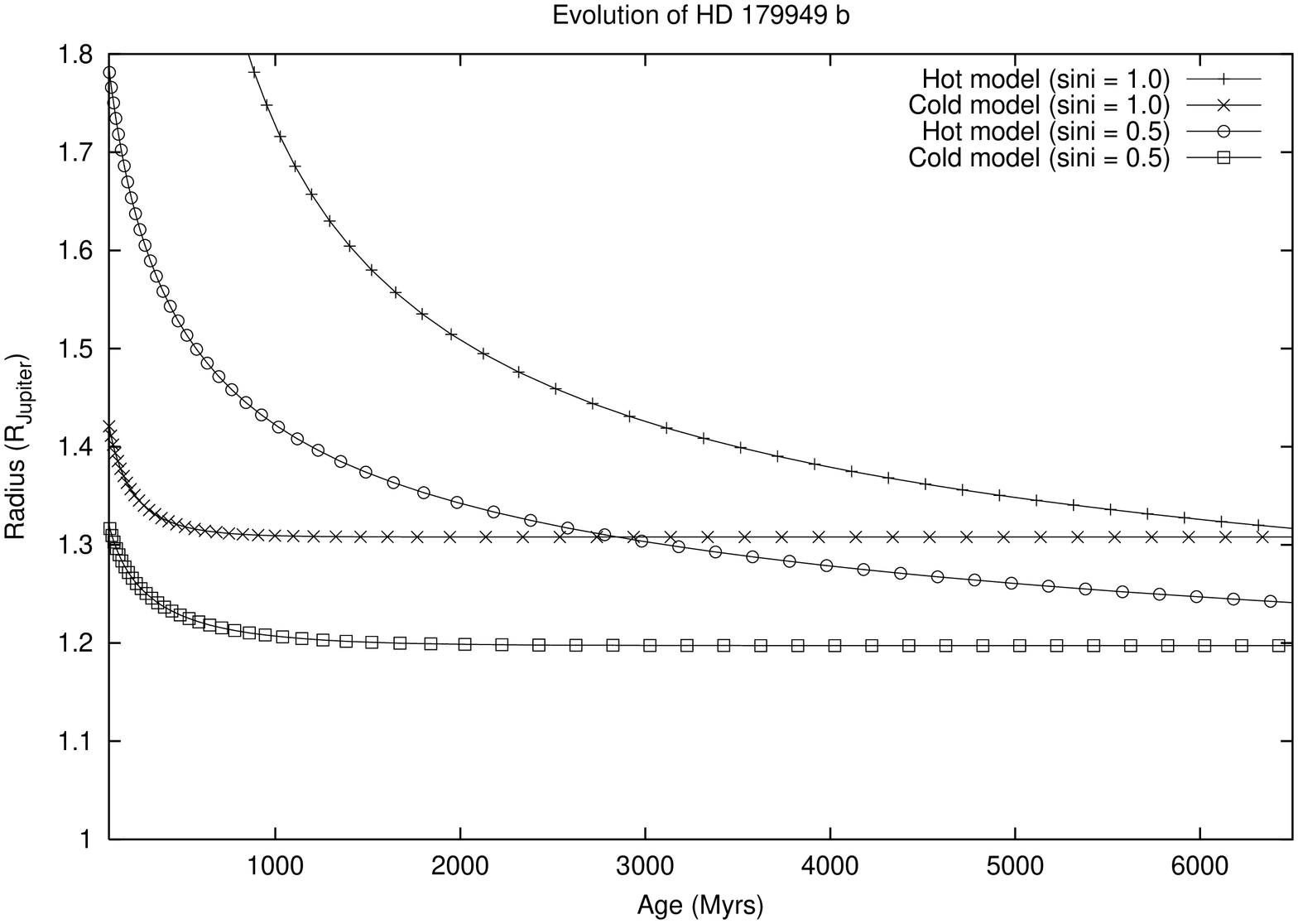,angle=-90,width=8cm} \\
\psfig{figure=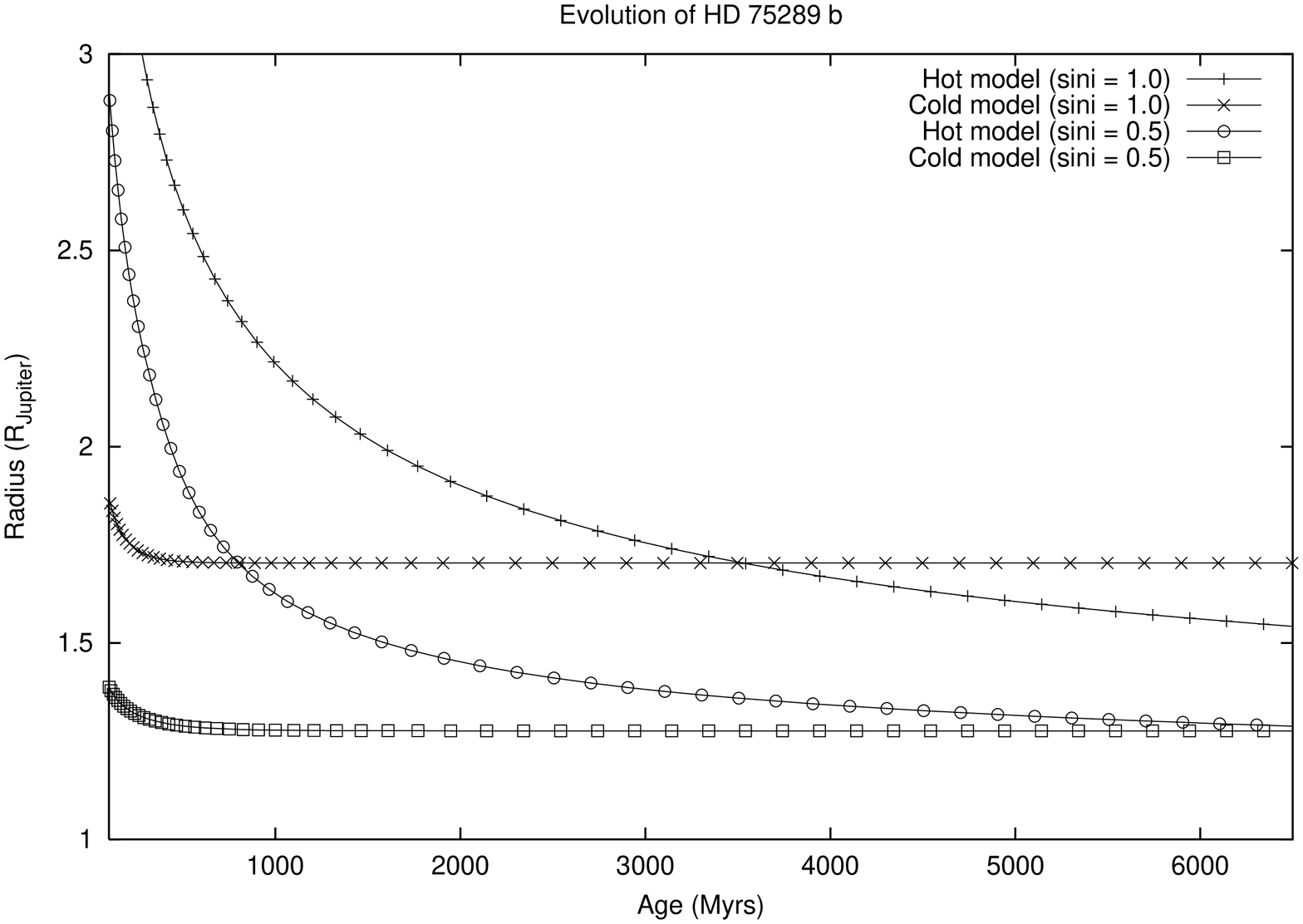,angle=-90,width=8cm}  &
\psfig{figure=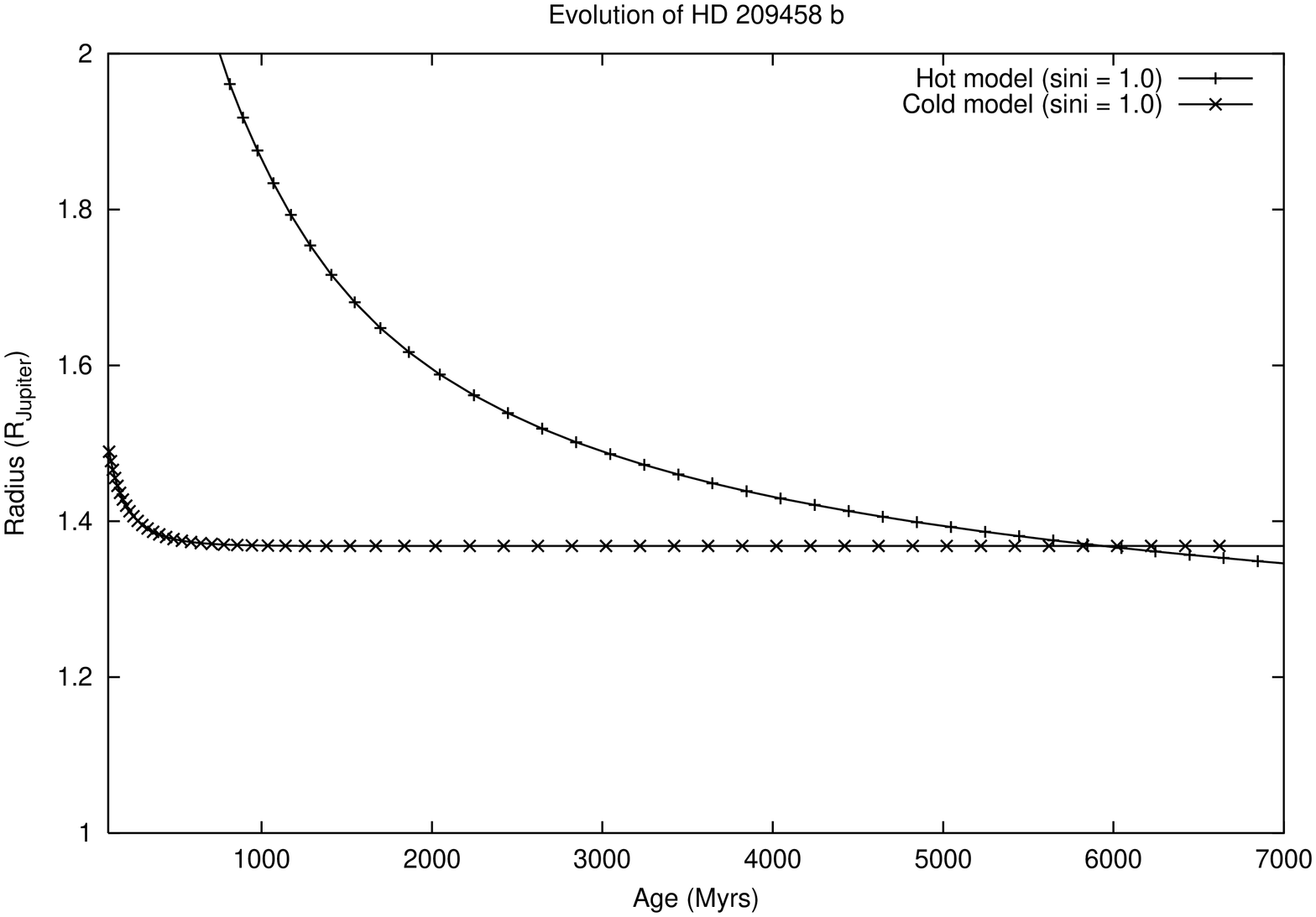,angle=-90,width=8cm} \\
\end{tabular}
\end{center}
\caption[]{Evolutionary radius models for the planets surrounding $\upsilon$ Andromeda,
$\tau$ Bootis, 51 Pegasi, HD 179949, HD 75289 and HD 209458. The evolutionary tracks
show both the ``Hot'' and ``Cold/Dissipative'' models for orbital inclinations of
$\sin i = 0.5$ and $\sin i = 1.0$ (Edge-on). Being a known transiting planet, HD 209458 b
is shown solely for $\sin i = 1.0$.
}
\label{fig:radius}
\end{figure*}

The known stellar parameters for 6 bright stars known (from Doppler wobble studies)
to harbour Pegasi planets are set out in Table~\ref{stellarparam}. Fits to the radial velocity data
have provided us with orbital periods for the planets, together with the
stellar reflex velocity $K_{*}$ of each star about the common centre-of-mass of the system.
Both parameters are set out in Table~\ref{planetparam} alongside calculated values for
the orbital radius of the system (from Kepler's Laws) and the minimum planet mass. In addition,
we provide theoretical estimates for the effective temperature and upper limit on the
radius of the planet (for edge-on orbits), assuming both ``Hot'' and ``Cold/Dissipative''
atmospheric structure models, as described in Section \ref{sec:radius}.

\subsection{Planetary Mass Estimates}
Among the known Pegasi planets there are several promising candidates for future 
spectroscopic reflected-light searches. The main selection criteria are that the host 
star must be bright, and that the planet/star flux ratio near superior conjunction 
must be high. Given spectroscopic and/or photometric estimates of the stellar
radius $R_{*}$, the projected rotation speed $v \sin{i}$ and rotation period
$P_{rot}$, we estimate
\begin{equation}
\sin{i} = \frac{P_{rot}~v\sin{i}}{2\pi R_{*}}
\label{eq:sini}
\end{equation}
This makes the assumption that the stellar rotation axis and the orbital plane
are close to perpendicular, as was determined for HD 209458 by \scite{queloz2000}.
The mass ratio $q=M_{p}/M_{*}$ follows from the observed orbital period $P_{orb}$
and stellar reflex velocity $K_{*}$:
\begin{equation}
\frac{q}{1+q} = \frac{K_{*} P_{orb}}{2\pi a \sin{i}}
\label{eq:massfunc}
\end{equation}
This then provides us with the planet's projected orbital velocity amplitude
$(K_{p}=K_{*}/q)$ in addition to the planet mass $(M_{p}=qM_{*})$.

\subsection{Planetary Radius Estimates}
\label{sec:radius}

Assuming a known mass, the radius of a gaseous planet is governed by
its Kelvin-Helmholtz contraction, and the following factors, in order
of significance: (1) its composition; (2) its atmospheric temperature;
(3) its age; and (4) possible additional energy sources. Uncertainties
also stem from our limited understanding of the underlying physics:
equations of state, opacities - see \scite{guillot99}.
 
The measurement of the radius of HD 209458 b \cite{charb2000,henry2000a,brown01}
showed that the planet is essentially made of hydrogen and
helium \cite{guillot96,burrows2000}, thus addressing the 
first source of uncertainty
concerning the radii of Pegasi planets. However, it was later shown
that the radius of HD 209458 b is too large to be explained by standard
evolution models using realistic atmospheric boundary conditions
\cite{boden01,guillot02,baraffe03}. Although \scite{burrows03}
question the need to invoke additional atmospheric energy sources,
\scite{boden01} proposed that HD 209458 b
may be kept inflated by the dissipation of tidal energy through the
circularization of its orbit. Due to the very small eccentricity of
the orbit (consistent with 0.0), \scite{showman02} proposed two
alternative explanations: either the atmosphere is relatively cold, as
predicted by radiative transfer calculations, and kinetic energy
generated in the atmosphere is dissipated relatively deep in the
interior (e.g. by tides due to a slightly asynchronous rotation of the
interior), or the atmosphere is hotter than calculated, for example
because shear instabilities force it into a quasi-adiabatic state.

\smallskip
\smallskip

\noindent We therefore calculate the radii of other Pegasi planets on the
following basis:
\begin{itemize}
\item All Pegasi planets have approximately the same composition (they
are mostly made of hydrogen and helium).
\item Their atmosphere is either ``hot'' or ``cold'', the temperature
at a given level (10 bars) being calculated as a function of the
effective temperature and gravity, as described in \scite{guillot02}
(see also \scite{burrows97}), assuming a bond albedo equal to 0.4.
\item In the ``cold'' case, an additional energy flux is deposited
at the centre of the planet. This flux is set equal to $8\times
10^{-3}$ times the absorbed stellar luminosity (i.e. between $1.2$ and
$3.1\times 10^{26}\rm\,erg\,s^{-1}$). This corresponds to the energy
flux necessary to reproduce the radius of HD 209458 b with the ``cold''
boundary conditions \cite{guillot02}.
\end{itemize}

Of course, a number of unknowns remain. Most importantly, both the
composition, bond albedo and the energy flux dissipated in the planet
are likely to be complex functions of the planetary mass, orbital
distance and even stellar type. However, this is likely to be a second
order effect because our radii calculations take advantage of the
known radius of HD 209458 b. It is interesting to see that even though
the parameters of the models have been set by the constraint on
HD 209458 b, the ``hot'' and ``cold'' scenarios shown in Figure
\ref{fig:radius} yield significantly different radii for the other
planets. This is due to the fact that energy dissipation (in the
``cold'' case) has a greater impact on planets of smaller mass, and
that the ``hot'' boundary condition yields a planetary radius that
is more sensitive to the orbital distance.

\subsection{Flux Ratio Estimates}
\label{sec:flux}

\begin{figure*}
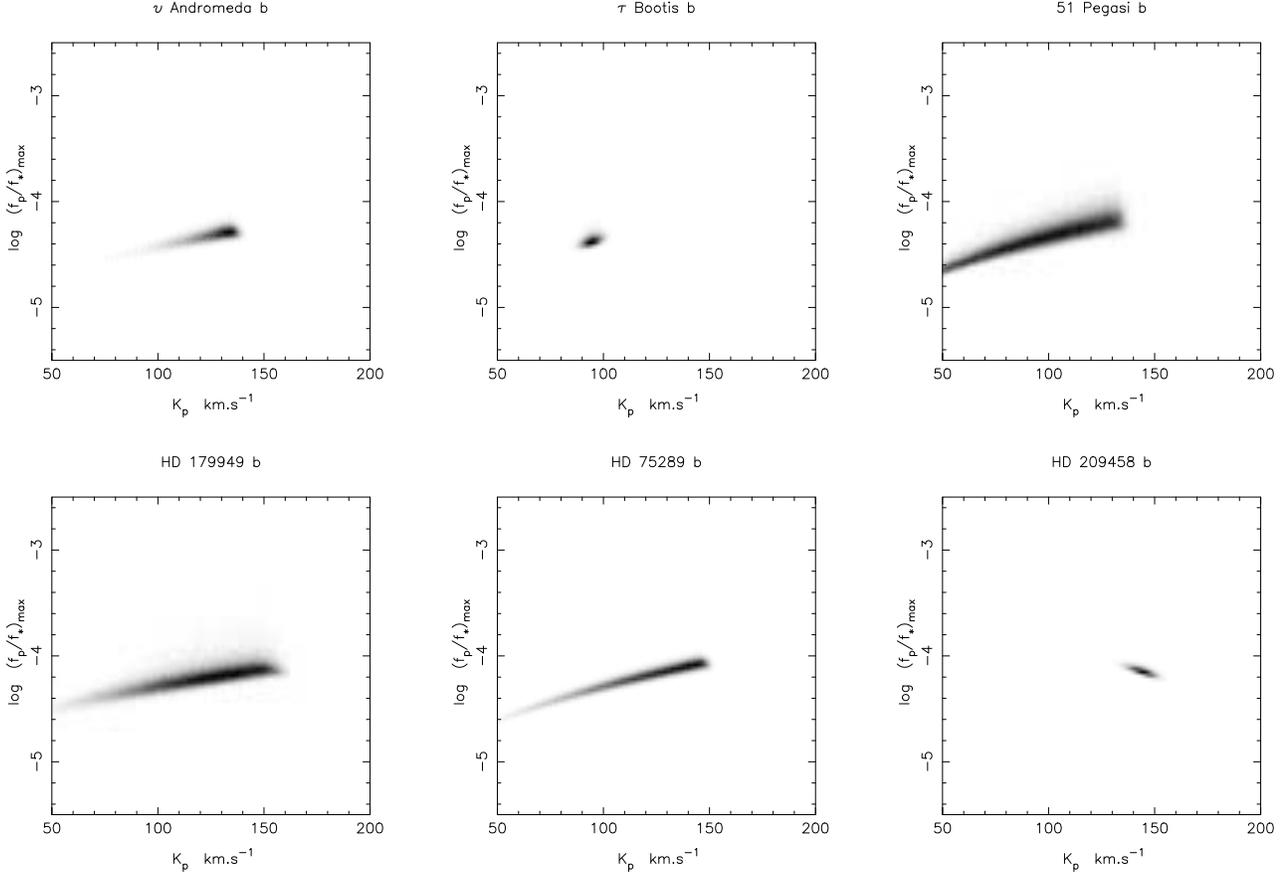

\begin{center}
\begin{tabular}{ccc}
\psfig{figure=md440fig2a.eps,bbllx=490pt,bblly=60pt,bburx=90pt,bbury=430pt,angle=-90,width=5.5cm} &
\psfig{figure=md440fig2b.eps,bbllx=490pt,bblly=60pt,bburx=90pt,bbury=430pt,angle=-90,width=5.5cm} &
\psfig{figure=md440fig2c.eps,bbllx=490pt,bblly=60pt,bburx=90pt,bbury=430pt,angle=-90,width=5.5cm} \\
\psfig{figure=md440fig2d.eps,bbllx=490pt,bblly=60pt,bburx=90pt,bbury=430pt,angle=-90,width=5.5cm} &
\psfig{figure=md440fig2e.eps,bbllx=490pt,bblly=60pt,bburx=90pt,bbury=430pt,angle=-90,width=5.5cm}  &
\psfig{figure=md440fig2f.eps,bbllx=490pt,bblly=60pt,bburx=90pt,bbury=430pt,angle=-90,width=5.5cm} \\
\end{tabular}
\end{center}
\caption[]{Prior probability density maps for the planet/star flux ratio near superior conjunction, as a function
of the planet's projected orbital velocity $K_{p}$, for an assumed geometric albedo $p=0.4$. The maps are
calculated for an evolutionary radius model that exhibits a ``Hot'' atmospheric structure, although maximum
likelihood values for both the ``Hot'' and ``Cold'' scenarios are shown at Table \ref{results}. Trials in which the
orbital inclination was high enough to cross the observer's line of sight to the stellar disc were rejected for all
cases bar HD 209458 b, since none of these other systems has been seen to exhibit transits
\cite{henry2000,tinney01}. For the transiting planet, its known inclination and mass act to constrain the
projected radial velocity amplitude.

\smallskip

\noindent In the case of $\tau Bootis$, an additional constraint was applied. Since the rotation of
the primary appears to have become synchronised with the planet's orbit
\cite{baliunas97,cameron99,henry2000}, the planet must have been massive enough to have synchronised
the star's spin within its own main sequence lifetime. For each trial we computed both the main
sequence lifetime and the synchronisation timescale. Trials in which the latter exceeded the former
were rejected. This, in conjunction with the well determined $v \sin{i}$ value, leads to a fairly
tight constraint on the system inclination and hence $K_{p}$.
}
\label{fig:prior}
\end{figure*}

\begin{table*}
\begin{tabular}{cccccc}
\hline \hline \\
Planet & $K_{p}$           & Inclination & $M_{p}/M_{Jup}$  & $(f_{p}/f_{*})_{max}$ & $(f_{p}/f_{*})_{max}$  \\
     & ($km.s^{-1}$)     & (Degrees)   &                  & Hot model             & Cold Model             \\
\hline \\
$\upsilon$ And b & 133  & 78 & 0.760 & 4.88 x $10^{-5}$ & 4.53 x $10^{-5}$   \\
$\tau$ Boo b     & 95   & 37 & 6.813 & 4.18 x $10^{-5}$ & 3.22 x $10^{-5}$   \\
51 Peg b         & 117  & 63 & 0.523 & 4.81 x $10^{-5}$ & 4.46 x $10^{-5}$   \\
HD 179949 b      & 135  & 60 & 0.945 & 6.44 x $10^{-5}$ & 5.34 x $10^{-5}$   \\
HD 75289 b       & 141  & 75 & 0.498 & 7.38 x $10^{-5}$ & 9.17 x $10^{-5}$   \\
HD 209458 b      & 144  & 87 & 0.620 & 7.18 x $10^{-5}$ & 7.17 x $10^{-5}$   \\
\hline \hline \\
\end{tabular}
\caption[]{Maximum likelihood statistics for the parameter distributions returned by the Monte
Carlo trials, given what we already know about the system from observation. In adopting a grey
albedo of $p = 0.4$ for our trials, similar to that of the Class V models of
\scite{sudarsky2000}. In reality, we would expect the albedo spectra of Pegasi planets to
exhibit a strong wavelength dependence. However, we believe the use of a grey albedo spectra
provides us with a realistic upper limit to the expected planet/star flux ratio.
}
\label{results}
\end{table*}

For a Lambert-sphere phase function $g(\alpha)$ we can compute the maximum expected
planet/star flux ratio near to superior conjunction,
\begin{equation}
(f_{p}/f_{*})_{max} = pg(\pi/2-i) \left( \frac{R_{p}}{a} \right) ^{2}.
\label{eq:fluxratio}
\end{equation}
We adopt the Lambert-sphere phase function because it provides a good approximation to the phase
functions of gas giants within our own solar system \cite{pollack86,charb99}. Although we expect the
albedo spectra of Pegasi planets to exhibit a strong wavelength dependence, here we adopt a simple
grey geometric albedo $p=0.4$ as a plausible comparison to the case of a high, reflective silicate
cloud deck with little overlying absorption, as in the Class V models of \scite{sudarsky2000}. Other
models in which cloud layers are deeper, or even absent, produce lower optical albedos; thus the flux
ratios we compute here should be treated as upper limits on the plausible planet/star flux ratio.

\section{Method Limitations}
\label{sec:pitfalls}

\begin{figure*}
\begin{center}
\begin{tabular}{cc}
\psfig{figure=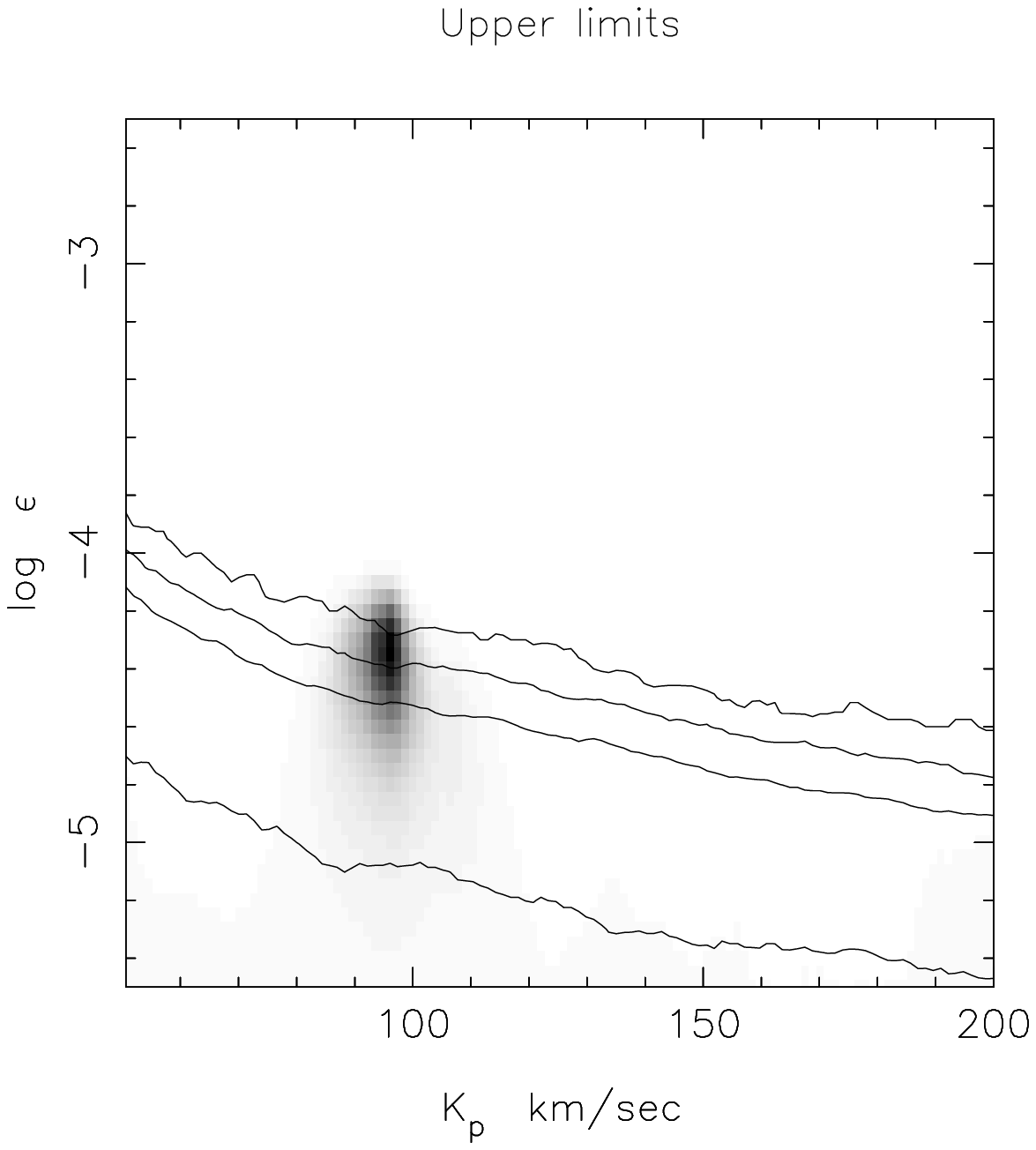,bbllx=490pt,bblly=60pt,bburx=90pt,bbury=430pt,angle=-90,width=5.5cm} &
\psfig{figure=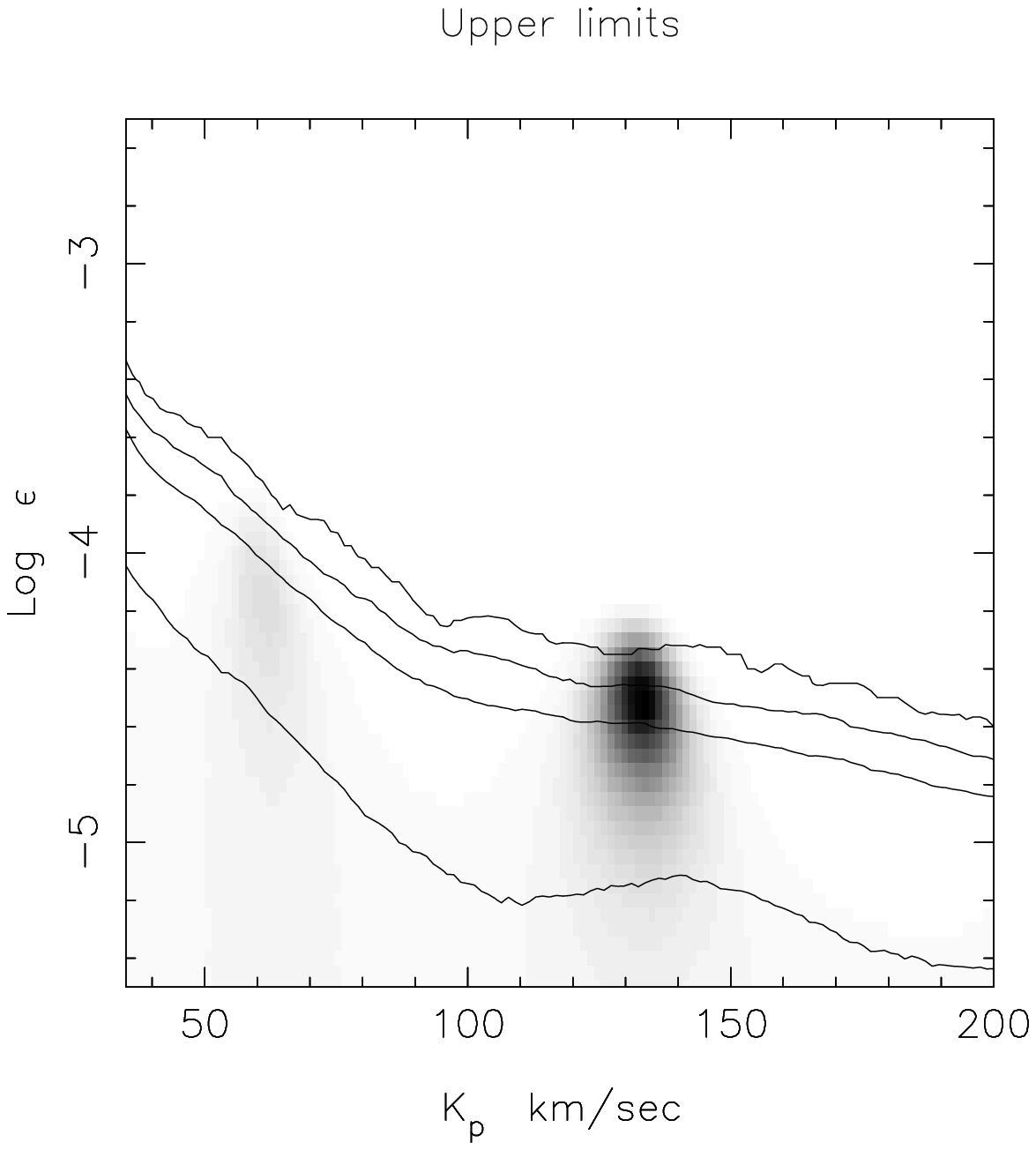,bbllx=490pt,bblly=60pt,bburx=90pt,bbury=430pt,angle=-90,width=5.5cm} \
\end{tabular}
\end{center}
\caption[]{Results from reflected-light studies of (a) $\tau$ Bootis \cite{leigh03} and (b)
$\upsilon$ Andromeda \cite{cameron02}, shown as relative probability maps of model parameters
$K_{p}$ and $\log(\epsilon_{0})=\log p(R_{p}/a)^{2}$. The results are derived from WHT/UES
observations of $\tau$ Bootis and $\upsilon$ Andromeda, assuming a grey albedo spectrum. The
greyscale denotes the probability relative to the best-fit model, increasing from 0 for white to
1 for black. The contours show the confidence levels at which candidate detections can be ruled
out as being caused by spurious alignments of non-Gaussian noise features. From top to bottom,
they show the 99.9\%, 99.0\%, 95.4\%\ and 68.3\%\ confidence limits. The dark features in each
probability map represent candidate planetary detections from the data but occur at too low a
confidence level to guarantee a genuine detection. We note, however, that in both cases the main
features appear close to the $K_{p}$ values predicted at Table \ref{results} by Monte Carlo
simulations and shown at Figure \ref{fig:prior}.
}
\label{fig:limits}
\end{figure*}
 
\begin{table*}
\begin{tabular}{ccccc}
\hline \hline \\
Target & $m_{V}$ & $(f_{p}/f_{*})_{max}$      & 4m-class   & 8m-class   \\
       & (Mags)  & Hot model                  & telescopes & telescopes \\
\hline \\
$\upsilon$ And & 4.09 & 4.88 x $10^{-5}$ & 20.0 hrs  & 5.0 hrs \\
$\tau$ Boo     & 4.50 & 4.18 x $10^{-5}$ & 39.8 hrs  & 9.9 hrs \\
51 Peg         & 5.46 & 4.81 x $10^{-5}$ & 72.7 hrs  & 18.2 hrs \\
HD 179949      & 6.25 & 6.44 x $10^{-5}$ & 84.0 hrs  & 21.0 hrs \\
HD 75289       & 6.35 & 7.38 x $10^{-5}$ & 70.1 hrs  & 17.5 hrs \\
HD 209458      & 7.65 & 7.18 x $10^{-5}$ & 245 hrs   & 61.3 hrs \\
\hline \hline \\
\end{tabular}
\caption[]{Exposure time predictor to compare benefits of observing different star and
telescope class combinations. We take 4m-class observations on $\upsilon$ Andromeda as
our standard, with an exposure time benchmark of 20.0 hrs required to reach the predicted
$(f_{p}/f_{*})_{max}$ signal levels. In deriving these estimates we assume that all
other variables are equal, such as overall instrument efficiency, observing strategies
and weather conditions.
}
\label{exposure}
\end{table*}

Spectroscopic reflected-light searches should provide us with a unique opportunity to
achieve the direct detection of a Pegasi planet, however there exist a number of
potential difficulties that require a cautious approach to any analysis.

(i) At the flux ratios we are working ($f_{p}/f_{*} < 10^{-4}$), we inevitably see
the appearence of systematic non-gaussian noise within data sets, noise that requires
to be carefully identified and corrected for in the subsequent analysis.

(ii) Any predictions from our searches require assumptions to be made about the nature of
the planetary system, such as the phase function $g(\alpha,\lambda)$, the planetary radius
$R_{p}$ or albedo spectra $p(\lambda)$, and that the orbital and stellar equatorial
planes are co-aligned. To some extent observational results will therefore act to
narrow the range of parameter space occupied by these Pegasi planets, although stronger
detections will allow us to draw more definite conclusions.

(iii) The ``Doppler Wobble'' method of detection confirms the existence of a
planet but cannot tell us its true orbital inclination. A spectroscopic
search for the reflected-light component of a close orbiting planet has the potential
to identify the projected orbital velocity $K_{p}$ and hence the orbital tilt.
However, with such a scope of possible $K_{p}$ values, it is important to estimate
where we might expect to detect the planet, given the known parameters of the system.
Such estimates, as detailed in Section \ref{sec:prob}, provide us with a quantifiable
method by which we can assess the plausibility of any candidate detection.

\section{Probability Distributions}
\label{sec:prob}

We derive the prior probability distributions for the observable quantities $K_{p}$ and
$(f_{p}/f_{*})_{max}$ by drawing randomly from the uncertainty distributions (assumed gaussian) 
for the observed quantities $K_{*}$, $M_{*}$, $R_{*}$, $P_{rot}$, Age and $v \sin{i}$. The
values of these quantities, for 6 of the brightest stars known to harbour Pegasi planets with orbital
periods of less than 5 days, are given in Tables~\ref{stellarparam} and \ref{planetparam}.
In order to provide a radius estimate for each planet $R_{p}$, we use our $\sin i$ estimate to
conduct a $\log{R_{p}}-\log{M_{p}}$ linear interpolation (for a selected Age) between the $\sin i = 1.0$
and $\sin i = 0.5$ evolutionary models. This is done for both the ``Hot'' and ``Cold'' evolutionary
scenarios. In each case we performed 1,000,000 random Monte Carlo trials to compute $K_{p}$ and
$\log (f_{p}/f_{*})_{max}$. The resulting joint probability distributions for the planet/star flux
ratio and projected orbital velocity amplitude are presented in Figure~\ref{fig:prior}.

\section{Discussion}

The maximum observable planet/star flux ratio for a planet depends strongly on the inclination of its
orbit to our line of sight, and on the distance of the planet from the star. The statistical 
analyses presented here provide estimates of the planet/star flux ratio, derived from the best data 
currently available on 6 bright stars harbouring planets with orbital periods of less 
than 5 days. These estimates are model-dependent, in that we have made theoretical 
assumptions about each planet's radius, the co-alignment of each orbital plane to the stellar
rotation axis and used a Lambert-sphere phase function to describe the atmospheric scattering
properties. In addition, we have adopted a grey geometric albedo $(p=0.4)$ that corresponds
roughly to the reflective, high-level silicate cloud decks predicted by the Class V spectral
models of \scite{sudarsky2000}.

Recent spectroscopic searches for relected-light signatures from $\tau$ Bootis b \cite{leigh03}
and $\upsilon$ Andromeda b \cite{cameron02}, shown at Figure \ref{fig:limits}, have already
reached detection limits comparable to the signal levels predicted here, using echelle
spectrographs on 4m-class telescopes. We find the total number of photons that must be
collected scales with the square of the planet/star flux ratio, whilst the length of time
required to collect this number of photons with a given telescope scales with the flux
received from the star. If we adopt 4m-class observations of $\upsilon$ Andromeda as our
benchmark (20.0 hrs) to reach the $(f_{p}/f_{*})_{max}$ signal levels predicted, then Table
\ref{exposure} details the relative exposure times for each planetary system,
given access to both 4m and 8m-class telescope facilities. Given that it has been possible to
probe to these deep signal levels for $\upsilon$ And in only a few nights on a 4m-class
telescope, the remaining 5 targets are clearly viable and compelling targets for future
studies with existing high-throughput spectrographs on 8m-class telescopes. Future
instruments that are also likely to be useful in this respect include the high-efficiency 
fibre-fed echelle spectrograph ESPADONS, which is to be commissioned on CFHT during 2003.

\bibliography{md440} \bibliographystyle{mn}

\begin{thebibliography}{{Goukenleuque, Bezard \& Lellouch}{1999}}

\bibitem[\protect\citefmt{Baliunas {\rm et~al.}}{1997}]{baliunas97}
Baliunas~S., Henry~G., Donahue~R., Fekel~F., Soon~W., 1997, ApJ, 474, 119

\bibitem[\protect\citefmt{Baraffe {\rm et~al.}}{2003}]{baraffe03}
Baraffe~I., Chabrier~G., Barman~T., Allard~F., Hauschildt~P., 2003, Astronomy
  and Astrophysics, 402, 701

\bibitem[\protect\citefmt{Barman, Hauschildt \& Allard}{2001}]{barman01}
Barman~T., Hauschildt~P., Allard~F., 2001, ApJ, 556, 885

\bibitem[\protect\citefmt{Barnes}{2001}]{barnes01}
Barnes~S., 2001, ApJ, 561, 1095

\bibitem[\protect\citefmt{Benz \& Mayor}{1984}]{benz84}
Benz~W., Mayor~M., 1984, Astronomy and Astrophysics, 138, 183

\bibitem[\protect\citefmt{Bodenheimer, Lin \& Mardling}{2001}]{boden01}
Bodenheimer~P., Lin~D., Mardling~R., 2001, ApJ, 548, 466

\bibitem[\protect\citefmt{Brown {\rm et~al.}}{2001}]{brown01}
Brown~T., Charbonneau~D., Gilliland~R., Noyes~R., Burrows~A., 2001, ApJ, 552,
  699

\bibitem[\protect\citefmt{Burrows {\rm et~al.}}{1997}]{burrows97}
Burrows~A., Marley~M., Hubbard~W., Sudarsky~D., Guillot~T., 1997, ApJ, 491, 856

\bibitem[\protect\citefmt{Burrows {\rm et~al.}}{2000}]{burrows2000}
Burrows~A., Guillot~T., Hubbard~W., M.~M., Saumon~D., Lunine~J., Sudarsky~D.,
  2000, ApJ, 534, 97

\bibitem[\protect\citefmt{Burrows, Sudarsky \& Hubbard}{2003}]{burrows03}
Burrows~A., Sudarsky~D., Hubbard~W., 2003, ApJ, In press : astro-ph/0305277

\bibitem[\protect\citefmt{Butler {\rm et~al.}}{1997}]{butler97}
Butler~P., Marcy~G., Williams~E., Hauser~H., Shirts~P., 1997, ApJ, 474, 115

\bibitem[\protect\citefmt{Charbonneau {\rm et~al.}}{1999}]{charb99}
Charbonneau~C., Noyes~D., Jha~S., Vogt~S., 1999, ApJ, 522, 145

\bibitem[\protect\citefmt{Charbonneau {\rm et~al.}}{2000}]{charb2000}
Charbonneau~D., Brown~T., Latham~D., Mayor~M., 2000, ApJ, 529, 45

\bibitem[\protect\citefmt{Cho {\rm et~al.}}{2003}]{cho03}
Cho~J., Menou~K., Hansen~B., Seager~S., 2003, ApJ, 587, 117

\bibitem[\protect\citefmt{Cody \& Sasselov}{2002}]{cody02}
Cody~A., Sasselov~D., 2002, ApJ, 569, 451

\bibitem[\protect\citefmt{Collier~Cameron {\rm et~al.}}{1999}]{cameron99}
Collier~Cameron~A., Horne~K., Penny~A., James~D., 1999, Nat, 402, 751

\bibitem[\protect\citefmt{Collier~Cameron {\rm et~al.}}{2002}]{cameron02}
Collier~Cameron~A., Horne~K., Penny~A., Leigh~C., 2002, MNRAS, 330, 187

\bibitem[\protect\citefmt{Fuhrmann, Pfeiffer \& Bernkopf}{1997}]{fuhrmann97}
Fuhrmann~K., Pfeiffer~J., Bernkopf~J., 1997, Astronomy and Astrophysics, 326,
  1081

\bibitem[\protect\citefmt{Fuhrmann, Pfeiffer \& Bernkopf}{1998}]{fuhrmann98}
Fuhrmann~K., Pfeiffer~J., Bernkopf~J., 1998, Astronomy and Astrophysics, 336,
  942

\bibitem[\protect\citefmt{Gonzalez \& Laws}{2000}]{gonzalez2000}
Gonzalez~G., Laws~C., 2000, AJ, 119, 390

\bibitem[\protect\citefmt{Gonzalez {\rm et~al.}}{2001}]{gonzalez01}
Gonzalez~G., Laws~C., Tyagi~S., Reddy~B., 2001, AJ, 121, 432

\bibitem[\protect\citefmt{Gonzalez}{1998}]{gonzalez98}
Gonzalez~G., 1998, Astronomy and Astrophysics, 334, 221

\bibitem[\protect\citefmt{Goukenleuque, Bezard \& Lellouch}{1999}]{gouken99}
Goukenleuque~C., Bezard~B., Lellouch~E., 1999, DPS (AAS) Conf., 31, 0906G

\bibitem[\protect\citefmt{Gratton, Focardi \& Bandiera}{1989}]{gratton89}
Gratton~R., Focardi~P., Bandiera~R., 1989, MNRAS, 237, 1085

\bibitem[\protect\citefmt{Gray, Napier \& Winkler}{2001}]{gray01}
Gray~R., Napier~M., Winkler~L., 2001, AJ, 121, 2148

\bibitem[\protect\citefmt{Gray}{1992}]{gray92}
Gray~D., 1992, The Observation and Analysis of Stellar Photospheres.
\newblock Cambridge University Press, Cambridge, U.K

\bibitem[\protect\citefmt{Groot, Piters \& van Paradijs}{1996}]{groot96}
Groot~P., Piters~A., van Paradijs~J., 1996, A\&AS, 118, 545

\bibitem[\protect\citefmt{Guillot \& Showman}{2002}]{guillot02}
Guillot~T., Showman~P., 2002, Astronomy and Astrophysics, 385, 156

\bibitem[\protect\citefmt{Guillot {\rm et~al.}}{1996}]{guillot96}
Guillot~T., Burrows~A., Hubbard~W., Lunine~J., Saumon~D., 1996, ApJ, 459, 35

\bibitem[\protect\citefmt{Guillot}{1999}]{guillot99}
Guillot~T., 1999, Sci, 286, 72

\bibitem[\protect\citefmt{Henry {\rm et~al.}}{2000a}]{henry2000}
Henry~G., Baliunas~S., Donahue~R., Fekel~F., Soon~W., 2000a, ApJ, 531, 415

\bibitem[\protect\citefmt{Henry {\rm et~al.}}{2000b}]{henry2000a}
Henry~G., Marcy~G., Butler~P., Vogt~S., 2000b, ApJ, 529, 41

\bibitem[\protect\citefmt{Houk \& Smith-Moore}{1988}]{houk88}
Houk~N., Smith-Moore~M., 1988, MSS, C04, 0H

\bibitem[\protect\citefmt{Leigh {\rm et~al.}}{2003}]{leigh03}
Leigh~C., Collier~Cameron~A., Horne~K., Penny~A., James~D., 2003, MNRAS, MNRAS
  accepted, astro-ph/0308413

\bibitem[\protect\citefmt{Marcy {\rm et~al.}}{1997}]{marcy97}
Marcy~G., Butler~P., Williams~E., Bildsten~L., Graham~J., Ghez~A., Jernigan~G.,
  1997, ApJ, 481, 926

\bibitem[\protect\citefmt{Marley, Gelino \& Stephens}{1999}]{marley99}
Marley~M., Gelino~C., Stephens~D., 1999, ApJ, 513, 879

\bibitem[\protect\citefmt{Mayor \& Queloz}{1995}]{mayor95}
Mayor~M., Queloz~D., 1995, Nat, 378, 355

\bibitem[\protect\citefmt{Mazeh {\rm et~al.}}{2000}]{mazeh2000}
Mazeh~T., Naef~D., Torres~G., Latham~D., Mayor~M., 2000, ApJ, 532, 55

\bibitem[\protect\citefmt{Pollack {\rm et~al.}}{1986}]{pollack86}
Pollack~J., Rages~K., Baines~K., Bergstrahl~J., 1986, Icarus, 65, 442

\bibitem[\protect\citefmt{Queloz {\rm et~al.}}{2000}]{queloz2000}
Queloz~D., Eggenberger~A., Mayor~M., Perrier~C., Beuzit~J., 2000, Astronomy and
  Astrophysics, 359, 13

\bibitem[\protect\citefmt{Seager \& Sasselov}{1998}]{seager98}
Seager~S., Sasselov~D., 1998, ApJ, 502, 157

\bibitem[\protect\citefmt{Showman \& Guillot}{2002}]{showman02}
Showman~P., Guillot~T., 2002, Astronomy and Astrophysics, 385, 166

\bibitem[\protect\citefmt{Sudarsky, Burrows \& Hubeny}{2003}]{sudarsky03}
Sudarsky~D., Burrows~A., Hubeny~I., 2003, ApJ, 588, 1121

\bibitem[\protect\citefmt{Sudarsky, Burrows \& Pinto}{2000}]{sudarsky2000}
Sudarsky~D., Burrows~A., Pinto~P., 2000, ApJ, 538, 885

\bibitem[\protect\citefmt{Tinney {\rm et~al.}}{2001}]{tinney01}
Tinney~C., Butler~R., Marcy~G., Jones~H., Penny~A., Vogt~S., 2001, ApJ, 551,
  507

\bibitem[\protect\citefmt{Udry {\rm et~al.}}{2000}]{udry2000}
Udry~S., Mayor~M., Naef~D., Pepe~F., Queloz~D., Santos~N., 2000, Astronomy and
  Astrophysics, 356, 590

\end{thebibliography}

\end{document}